# Multilayer graphene under vertical electric field

S. Bala kumar and Jing Guo[a]

*Department of Electrical and Computer Engineering, University of Florida, Gainesville, Florida 32608, USA*

Abstract

We study the effect of vertical electric field (E-field) on the electronic properties of multilayer graphene. We show that the effective mass, electron velocity and density-of-state of a bilayer graphene are modified under the E-field. We also study the transformation of the band structure of multilayer graphenes. E-field induces finite (zero) bandgap in the even (odd)-layer ABA-stacking graphene. On the other hand, finite bandgap is induced in all ABC-stacking graphene. We also identify the optimum E-field to obtain the maximum bandgap in the multilayer graphenes. Finally we compare our results with the experimental results of a field-effect-transistor.

---

[a] Corresponding author. E-mail: guoj@ufl.edu

The research activities in the graphene based materials show a considerable growth due to the remarkable physical properties of graphene[1,2]. Graphene has a very large intrinsic carrier mobility[3], which makes it a very promising material for device application. However, a prestine graphene is semimetal with zero bandgap[1], and thus the creation of bandgap is important to enable the application of graphene in electronic[4] and photonics devices[5,6]. One of the methods to induce bandgap in graphene is by breaking the inversion symmetri[7-12] via the application of vertical electric field in bilayer graphenes. When a two graphene layers are stacked in the third dimension, the interlayer coupling modifies the electronic structure of the graphene layers. Moreover, vertical electric field ($E^\perp$-field) applied perpendicular to the graphene layers would create an interlayer potential asymmetric, which induces bandgap. This $E^\perp$-field induced gap, in the bilayer graphene has been demonstrated theoretically[7-10] and experimentally[11-14].

In this letter, we study the effect of $E^\perp$-field on the electronic structure of multilayer graphenes. Generally, there are two typical ways to stack the graphene layers, i.e. ABA-stacking and ABC-stacking[15]. We refer ABA-stacking (ABC-stacking) graphene as ABA-graphene (ABC-grapnene). The stability, band structure and their dependence on the $E^\perp$-field of the multilayer graphenes have been examined by M. Okaki et al.[9] using density function theory. M. Koshino[10] has further studied the interlayer screening effect in the multilayer graphenes. Here, we first examine the modification in the electronic structure of a bilayer graphene (AB-stacking) due to the $E^\perp$-field. We derive the expressions for the effective mass and electron velocity of the conduction band, to study the influence of $E^\perp$-field on these parameters. We further extend the study to more number of layers. Even-layer (odd-layer) ABA-graphenes show parabolic (both linear and parabolic) dispersions, while the ABC-graphanes has only parabolic dispersions. Depending on the stacking configuration and the number of layers, electric field may induce a finite bandgap in the graphene layers. Under $E^\perp$-field, a finite bandgap is induced in even-layer ABA-graphenes while odd-layer ABA-graphenes remain metallic. On the other hand, a finite gap is induced in all ABC-graphene layers. Finally, we simulate the variation of bandgap of a graphene channel in a field effect transistor (FET) structure and compare our results to the experimental results.

The π-orbital tight binding (TB) Hamiltonian of an AB-stacking graphene is given by,

$$H(k) = \begin{bmatrix} U & \lambda(K_x,K_y) & \gamma(K_x,K_y) & \eta(K_x,K_y) \\ \lambda(K_x,K_y)^* & U & t_\perp & \gamma(K_x,K_y) \\ \gamma(K_x,K_y)^* & t_\perp & -U & \lambda(K_x,K_y) \\ \eta(K_x,K_y)^* & \gamma(K_x,K_y)^* & \lambda(K_x,K_y)^* & -U \end{bmatrix} \quad (1)$$

,where $\lambda(K_x,K_y) = t_0\left(1+e^{iK_1}+e^{iK_2}\right)$, $\gamma(K_x,K_y) = t_\perp\left(1+e^{iK_1}+e^{iK_2}\right)$,

$\eta(K_x,K_y) = t_\angle\left(e^{i3K_x}+e^{iK_1}+e^{iK_2}\right)$, $K_{1(2)} = \left(3K_x+(-)\sqrt{3}K_y\right)/2$, $K_{x(y)} = a_{cc}k_{x(y)}$, and the intralayer atomic distance $a_{cc}$=0.142nm. Equation (1) is derived by modifying the nearest-neighbor (NN) TB Hamiltonian[8,10,16]. In addition to the NN-interactions, here, we also consider the second-NN-interlayer (SNNI) interaction. The NN-interlayer (NN-intralayer) hopping parameter is $t_o$=2.7eV ($t_\perp$=0.3eV). The SNNI hopping parameter is $t_\angle$=0.14eV. The $E^\perp$-field across the bilayer graphene is given by, $F = 2U/d/q$, where d=0.335nm is the interlayer distance, and electron charge $q = 1.602\times10^{-19}$. For analytical discussions, we neglect the SNNI [refer Fig. 2], and thus the lowest conduction band (CB) of the bilayer graphene is

$$E_C(K_x,K_y) = \sqrt{U^2 + |\lambda(K_x,K_y)|^2 + t_\perp^2/2 - \sqrt{|\lambda(K_x,K_y)|^2\left(4U^2+t_\perp^2\right)+t_\perp^4/4}} \quad (2)$$

The bandgap occurs when $E_C$ is minimum i.e.

$$\frac{d(E^2)}{d\lambda} = 0 \Rightarrow \lambda_{gap}(U) = \sqrt{\frac{2U^2(2U^2 + t_\perp^2)}{4U^2 + t_\perp^2}} \tag{3}$$

Replacing (3) into (2), we can express the bandgap, $E_{gap} = 2E_{C,\min}$ as $E_{gap}(U) = \sqrt{\frac{4t_\perp^2}{t_\perp^2/U^2 + 4}}$. This sets the maximum limit of the $E_g$ to $t_\perp$. The black region in Fig. 1(b), indicates the bandgap, which increases from zero to a maximum value as the $E^\perp$-field increases.

As the $E^\perp$-field increases there is also a significant change in the band structure. In graphene, one of the valleys where bandgap occurs is located along $K_y = \frac{4\pi}{3\sqrt{3}} \equiv K_{y,gap}$, and thus we define

$\left|\lambda(K_x, K_{y,gap})\right| = \left|2t_0 \sin\frac{3K_x}{4}\right| \equiv \lambda_0(K_x)$. Equating $\lambda_0(K_x) = \lambda_{gap}(U)$, we get the $K_x$ at the bandgap as a function of electric field, i.e. $|K_{x,gap}| = \frac{4}{3}\sin^{-1}\left(\frac{\lambda_{gap}(U)}{2t_0}\right)$. As shown in Fig.1 (a), as the $E^\perp$-field increases the bandgap occurs at a higher $|K_x|$. The change in the band structure also gives rise to interesting modifications of the electron velocity, density of state, and effective mass of the CB. The effective mass of the conduction electron at $K_x=0$, $m_0^*$ and at the bandgap, $m_{gap}^*$ as well as the electron velocity, v are

$$m_0^* = -\frac{\hbar^2}{a_{cc}^2} \frac{t_\perp^2}{9t_0^2 U}, \qquad \text{for } U > 0 \tag{4a}$$

$$m_{gap}^* = \frac{\hbar^2}{a_{cc}^2} \frac{t_\perp(4U^2 + t_\perp^2)^{5/2}}{9U(2U^2 + t_\perp^2)(8U^2 t_0^2 + 2t_0^2 t_\perp^2 - U^2 t_\perp^2 - 2U^4)}, \qquad \text{for } U > 0 \tag{4b}$$

$$v(K_x) = \frac{a_{cc}}{\hbar} \frac{3t_o^2 \left(2\beta(K_x) - 4U^2 - t_\perp^2\right)}{2\beta(K_x)\sqrt{4U^2 + 2t_\perp^2 + 4\lambda_o(K_x)^2 - 4\beta(K_x)}} \sin\left(\frac{3K_x}{2}\right) \quad (4c)$$

, where $\beta(K_x) = \sqrt{\lambda_o(K_x)^2 \left(4U^2 + t_\perp^2\right) + t_\perp^4/4}$. For bilayer graphene, when F=0 the $K_{x,gap}$ =0, and the CB has a finite effective mass, i.e. $m_0 = m_{gap} = 0.035 m_e$. As the $E^\perp$-field increases, $K_{x,gap}$ increases while the curvature of the band becomes negative (positive) at $K_x$=0 ($K_x$= $K_{x,gap}$). When a small $E^\perp$-field is applied, the $m_0^* \to +\infty$ and $m_{gap}^* \to -\infty$. Referring to Fig. 1a(i), at the bandgap the $m_{gap}^*$ decreases drastically to a finite positive value with increasing $E^\perp$-field. At $K_x$=0, as the $E^\perp$-field increases, the curvature gets sharper, and the difference between the first and the second conduction band, $\Delta$ approaches zeros. As a result, at large $E^\perp$-field, the dispersion curve at the vicinity of $K_x$=0 becomes linear with zero effective mass (massless fermion), i.e. $m_0^* = 0$. The Fig. 1a(ii) shows the electron velocity at positive $K_x$ values. Due to the formation of negative curvature at $K_x$=0, the electron velocity of a positive $K_x$ switches from positive to negative as the field increases. The sign of the velocity changes when $K_{x,gap} > K_x$. Therefore as shown in Fig. 1a(ii), for larger $K_x$, the transition occurs at larger $E^\perp$-field.

Next we study the effect of $E^\perp$-field on the multilayer graphene, with the total number of layer N. The results are shown in Fig. 2 and 3, where the screening effect[8,10,11,17] is taken into account. Due to the screening effect, when an $E^\perp$-field of $F_{app}$ is applied across the multilayer graphene, the effective $E^\perp$-field is reduced to $F_{eff}$. We use a self-consistent method to compute the screened effect[10].

The π-orbital tight binding Hamiltonian for ABA-graphene (ABC-graphene) is

$$H_{ABA(ABC)} = \begin{bmatrix} H_1(H_1) & \gamma(\gamma) & & \\ \gamma^\dagger(\gamma^\dagger) & H_2(H_2) & \gamma^\dagger(\gamma) & \\ & \gamma(\gamma^\dagger) & H_3(H_3) & \gamma(\gamma) \\ & & \gamma^\dagger(\gamma^\dagger) & \ddots & \ddots \\ & & & \ddots & \ddots \end{bmatrix} \quad (5)$$

, where $H_j = \begin{bmatrix} U_j & \lambda(K_x,K_y) \\ \lambda(K_x,K_y)^* & U_j \end{bmatrix}$, $\gamma = \begin{bmatrix} \gamma(K_x,K_y) & \eta(K_x,K_y) \\ t_\perp & \gamma(K_x,K_y) \end{bmatrix}$ and $U_j$ is the electrochemical potential at the j-th layer.

First, we consider the ABA-graphene. Neglecting the SINN, when U=0, the eigenenergies of the N-layer graphene is given such that

$$E(K_x,K_y)^2 = |\lambda(K_x,K_y)|^2 + 2t_\perp^2 C_\alpha^2 \pm 2t_\perp C_\alpha \sqrt{|\lambda(K_x,K_y)|^2 + t_\perp^2 C_\alpha^2} \quad (6)$$

, where $C_\alpha = \left|\cos\dfrac{\alpha\pi}{N+1}\right|$. The α refers to the index of the subbands. For N-layer graphene, α varies from 1 to 2N, i.e α=1,2,…,2N. When $C_\alpha = 0$ ($C_\alpha > 0$), the energy band is linear (approximately parabolic) similar to the monolayer (bilayer) graphene. In odd-layer graphene, $C_\alpha = 0$ when α=(N+1)/2, while in even-layer graphene, $C_\alpha > 0$ for all the bands. This is shown in Fig. 2(a), where near $K_x$=0, a linear (approximately parabolic) energy dispersion is obtained for odd (even) N. For the even-layer graphene, as the N increases, $C_{\alpha=(N+1)/2} \to 0$, and thus the dispersion approaches linearity identical the monolayer graphene.

Fig. 2(b) and (c) show the effect of $E^\perp$-field on the lowest (highest) conduction (valence) sub-band of the ABA-graphene. The $E^\perp$-field induces a finite bandgap for the even-layer graphenes. On the other hand, in the odd-layer graphenes the lowest conduction and the highest valence band touches at the dirac point, resulting in zero bandgap, independent of the applied $E^\perp$-field. It can be noted that the effect of $E^\perp$-field induced bandgap on the odd (even) layer graphene is qualitatively identical to the monolayer (bilayer) graphene. Referring to Fig. 2(d) and (e), for the even-layer graphenes, only the bilayer graphene shows a significant bandgap at higher $E^\perp$-field. When N is larger, the magnitude of the induced bandgap decreases considerably. For the ABA-stacking with large stacks of N>4, the bandgap remains approximately zero independent of the $E^\perp$-field. Therefore, in a thicker ABA-stacking graphene the transport properties are not significantly modified by the E-field.

Next we study the effect in ABC-graphene. Referring to Fig. 3(a), at zero $E^\perp$-field, an approximately parabolic dispersion is obtained for all the values of N. When N increases, the curvature of dispersion at the dirac point ($K_x$=0) decreases, indicating heavier electron/hole at the dirac point. Unlike ABA-graphene, in the ABC-graphene, bandgap is induced for both odd and even N (except N=1). In general, for the ABC-graphene the bandgap increases to a maximum value, and then it decreases. The maximum bandgap, is the largest in the bilayer graphene, and decreases as the N increases. However, the optimum $E^\perp$-field where the bandgap is maximum, can be reduced by increasing the number of layer. Therefore, for small $E^\perp$-field, ABC-graphenes with N>2 may have larger bandgap compared to the bilayer graphenes. For example, as shown in Fig. 3(d), for E-field of $F_{app}$ = 0.5eV/nm, the maximum bandgap is obtained when the number of layers, N=3.

One of the important applications of the $E^\perp$-field induced bandgap is the use of graphene as a channel in FETs. The increase of bandgap at higher field would reduce the leakage current at the OFF-state, resulting in a higher ON-OFF current ratio. Therefore we study the effect of $E^\perp$-field on the bandgap in a dual gated FET device. For the simulation, we model a graphene based FET device as shown in ref. [13]. Figure 4(a) shows the structure of the device. Ref [13] shows that for the graphene layers, the maximum channel resistance changes as $U_b$ is varied from -10eV to 30eV. The change in the maximum resistance is attributed to the variation of bandgap due to applied $E^\perp$-field. In Fig. 4(a), we show the variation of the bandgap as a function of average displacement field, $D_{ave}$. The displacement field is defined as $D_{ave} = (D_b+D_t)/2$, where $D_b = +\varepsilon_b(U_b-U_b^0)/d_b$ and $D_t = +\varepsilon_t(U_t-U_t^0)/d_t$. Here $\varepsilon$, d, and $U^0$ represents the dielectric constant, the thickness of the dielectric layers and Dirac offset potential due to the initially environmentally induced carrier doping, respectively. The values of $\varepsilon$ and d are shown in Fig. 4(a), while the values for $U_{t,b}$ and $U_{t,b}^0$ are extracted from the results in ref. [13]. Referring to Fig. 4(b), the simulation results for the AB-stacking graphene is in a close qualitative agreement with the experimental results (marked by '+'). However quantitatively our simulation shows a bandgap opening of almost two times larger than the experimental results. The experimentally measured effective gap is smaller than the actual bandgap due to the additional channels for flow created by the disorders.[14] It is worth noting, that, at similar bias conditions, our results are quantitatively similar to the results obtained through the optical measurements [5,11] (marked by 'x' in Fig. 4(b)). Next we replace the channel with the trilayer graphene. Trilayer graphene with ABA-stacking does not show any bandgap opening, while the ABC-stacking predicts a bandgap opening which is much larger than that obtained in the experimental results, as shown in Fig. 4(c). Therefore, the third layer in the experiment is most likely disoriented, such that the multilayer acts as a bilayer with the third, largely uncoupled, layer partially screening the E-field.

The work was supported by NSF, ONR, and ARL.

References

[1] Ph. Avouris, Z. Chen, V. Perebeinos, *Nat. Nanotechnol. 2*, 605 (**2007**); A. H. C. Neto, F. Guinea, N. M. R. Peres, K. S. Novoselov, A. K. Geim, *Rev. Mod. Phys. 81*, 109 (**2009**).

[2] R. R. Nair, P. Blake, A. N. Grigorenko, K. S. Novoselov, T. J. Booth, T. Stauber, N. M. R. Peres, A. K. Geim, *Science 320*, 1308 (**2008**); C. Chen, S. Rosenblatt, K. I. Bolotin, W. Kalb, P. Kim, I. Kymissis, H. L. Stormer, T. F. Heinz, J. Hone, J. *Nat. Nanotechnol. 4*, 861 (**2009**).

[3] K. I. Bolotin, K. J. Sikes, Z. Jiang, D. M. Klimac, G. Fudenberg, J. Hone, P. Kim, H. L. Stormer, *Solid State Commun. 146*, 351 (**2008**); X. Du, I. Skachko, A. Barker, E. Y. Andrei, *Nat. Nanotechnol. 3*, 491 (**2008**).

[4] P. San-Jose, E. Prada, E. McCann, H. Schomerus, *Phys. Rev. Lett. 102* 247204 (**2009**); [8] M.Tonouchi, *Nature Photon. 1*, 97 (**2009**).

[5] Y. Zhang, T. Tang, C. Girit, Z. Hao, M. C. Martin, A. Zettl, M. F. Crommie, Y. R. Shen, F. Wang, *Nature 459*, 820 (**2009**).

[6] F. Wang, Y. Zhang, H. Tian, C. Girit, C.; Zettl, A.; Crommie, M.; Shen, Y. R. *Science 320*, 206 (**2008**)

[7] E. McCann and V. I. Fal'ko, Phys. Rev. Lett. 96, 086805 (**2006**); H. Min, B. R. Sahu, S. K. Banerjee, and A. H. MacDonald, Phys. Rev. B **75**, 155115 (**2007**).

[8] McCann, E. *Phys. Rev. B 74*, 161403(R) (**2006**).

[9] M. Aoki and H. Amawashi, Solid State Commun. **142**, 123 (**2007**)

[10] M. Koshino, Phys. Rev. B **81**, 125304 (**2010**).

[11] K. F. Mak, C. H. Lui, J. Shan, T. F. Heinz, *Phys. Rev. Lett. 102*, 256405 (**2009**).

[12] A. B. Kuzmenko, I. Crassee, D. van der Marel, P. Blake, K. S. Novoselov, *Phys. Rev. B 80*, 165406 (**2009**); E. V. Castro, K. S. Novoselov, S. V. Morozov, N. M. R. Peres, J. M. B. Lopes dos Santos, J. Nilsson, F. Guinea, A. K. Geim, and A. H. Castro Neto, Phys. Rev. Lett. **99**, 216802 (**2007**); J. B. Oostinga, H. B. Heersche, X. Liu, A. F. Morpurgo, and L. M. K. Vandersypen, Nature Mater. **7**, 151 (**2008**).

[13] W. Zhu, D. Neumayer, V. Perebeinos, and P. Avouris, Nano Lett., 10, 3572 (**2010**)

[14] F. Xia, D. B. Farmer, Y. Lin, Ph. Avouris, *Nano Lett. 10*, 715 (**2010**).

[15] H. Lipson and A. R. Stokes, Proc. R. Soc. London **A181**, 101 (**1942**).

[16] F. Guinea, A. H. Castro Neto, and N. M. R. Peres, Phys. Rev. B **73**, 245426 (2006).

[17] M. Koshino and E. McCann, Phys. Rev. B **79**, 125443 (**2009**).

Captions

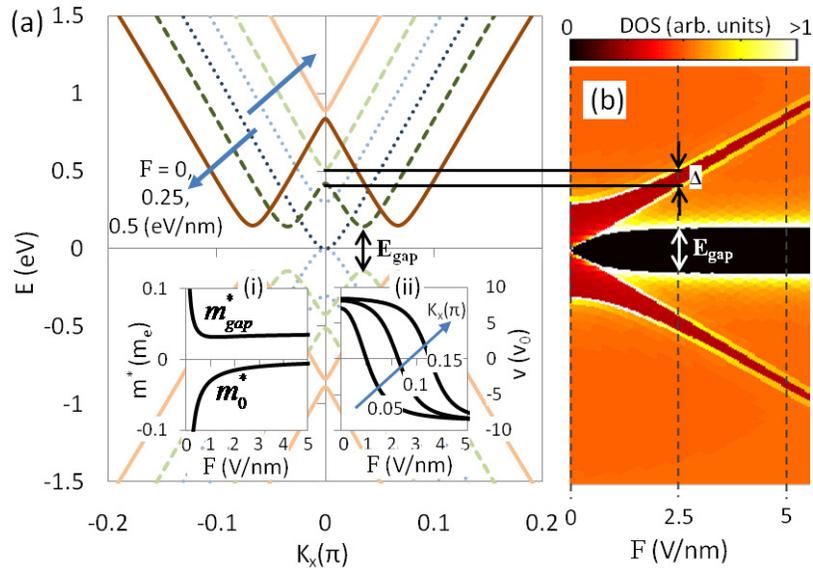

Figure 1 (a) Shows the band diagram modification of a bilayer graphene along $K_y=K_{y,min}$ when the vertical E-field is 0, 0.25, and 0.5V/nm. (b) Color map showing the DOS with increasing E-field. The black region indicated the bandgap. The difference between the first and the second conduction band is labeled by $\Delta$. Inset (i) shows the effective mass variation. $m_0$ ($m_{gap}$) refers to the effective mass at Kx=0 (the bandgap). Inset (ii) shows the conduction electron velocity at a positive $K_x$ value as a function of $F$. In all the above results, the SINN was neglected. [$v_0=10^5$m/s, electron rest mass, $m_e=9.109\times10^{-31}$kg]

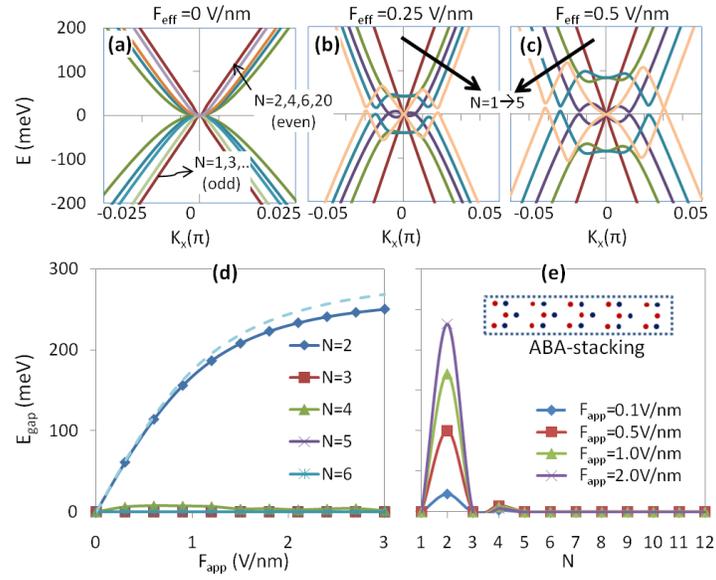

Figure 2 The band structure of multilayer ABA-stacking graphene, when the effective E-field (a)$F_{eff}$=0, (b) $F_{eff}$=0.25V/nm, and (c)$F_{eff}$=0.5V/nm. Only the lowest (highest) conduction (valence) is shown. (d) Bandgap as a function of the applied E-field ($F_{app}$) for the ABA-stacking graphene. (e) Bandgap as a function of number of layers for the ABA-stacking graphene. SINN is included in all the results. The dotted line (d) compares the result without SINN for N=2.

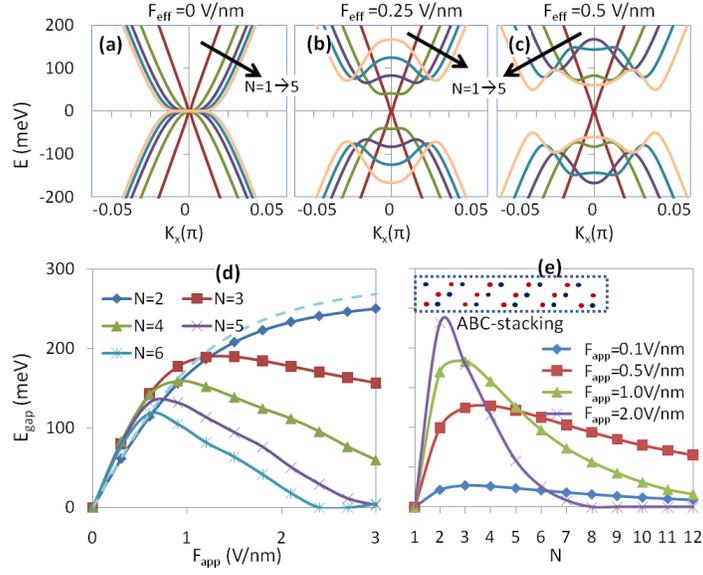

Figure 3 The band structure of multilayer ABC-stacking graphene, when the effective E-field (a)$F_{eff}$=0, (b) $F_{eff}$=0.25V/nm, and (c)$F_{eff}$=0.5V/nm. Only the lowest (highest) conduction (valence) is shown. (d) Bandgap as a function of the applied E-field ($F_{app}$) for the ABA-stacking graphene. (e) Bandgap as a function of number of layers for the ABA-stacking graphene. SINN is included in all the results. The dotted line (d) compares the result without SINN for N=2.

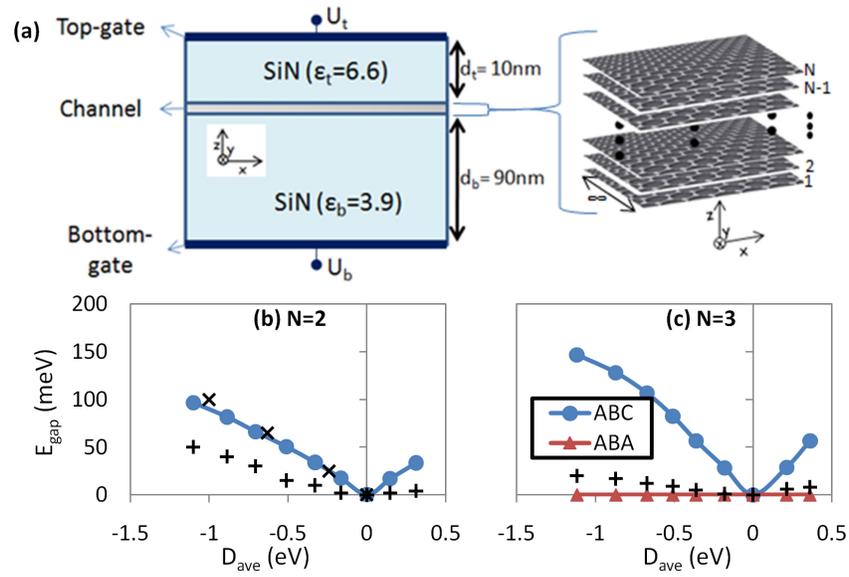

Figure 4 (a) Schematic diagram of the FET device used in the simulations. The channel is made of graphene layers. All diagrams are not drawn in scale. The bandgap, $E_{gap}$ dependence on the average displacement field, $D_{ave}$ are shown for (b) number of layers, m=2 and (c) m=3. For (c) the results for two type of layer stacking, i.e. ABA- and ABC- stacking are shown. '+' indicates the experimental gap [see ref. 13]. 'x' indicates the gap obtained in the optical measurements [see ref. 11].